\documentclass[aps,pra,groupedaddress,showpacs]{revtex4}

\begin{document}


\title{$\frac{1}{N}$-expansion for the Dicke model and the decoherence program}


\author{Marco Frasca}
\email[e-mail:]{marcofrasca@mclink.it}
\affiliation{Via Erasmo Gattamelata, 3 \\
             00176 Roma (Italy)}


\date{\today}

\begin{abstract}
An analysis of the Dicke model, $N$ two-level atoms interacting with a single
radiation mode, is done using the Holstein-Primakoff transformation. The main aim
of the paper is to show that, changing the quantization axis with respect to the
common usage, it is possible to
prove a general result either for $N$ or 
the coupling constant going to infinity for the exact solution of the model. 
This completes the analysis, known in the current literature, with
respect to the same model in the limit of $N$ and volume going to infinity, keeping the density constant.
For the latter the proper axis of quantization is given by the Hamiltonian of the
two-level atoms and for the former the proper axis of quantization is defined by
the interaction.
The relevance of this result relies on the observation that a general
measurement apparatus acts using electromagnetic interaction and so, one can states
that the thermodynamic limit is enough to grant the appearance of classical effects.
Indeed, recent experimental results give first evidence
that superposition states disappear interacting with an electromagnetic field having
a large number of photons. 
\end{abstract}


\maketitle


\section{Introduction}

One of the fundamental questions that a possible interpretation of quantum
mechanics needs to answer is where to put the boundary between the quantum and
the classical world. This question lies at the heart of the problem of
measurement. A possible answer to this question is given by environmental 
decoherence \cite{zur1} where it is assumed that the disappearance of entangled
states between a quantum system and the measurement apparatus is due to the effect
of an external environment having effects even more difficult to eliminate as the
measurement apparatus becomes larger, being openness of systems rather general. 
This approach cannot solve the problem of measurement in quantum mechanics \cite{crit} 
but gives a proper direction for its solution.

Recent experiments realized by Haroche et al. \cite{har1,har2} have given the first
sound evidence that classicality can emerge also without requiring any environment.
They were able to realize experimentally the asymptotic states with a large
number of photons firstly devised by Gea-Banacloche \cite{gb1,gb2} and further
studied by Knight et al. \cite{kn1,kn2}. Gea-Banacloche put forward the idea that
the measurement process could be effectively represented by these states proper to
the Jaynes-Cummings model \cite{sch}, a rather ubiquitous Hamiltonian describing
radiation-matter interaction. The analysis was not enough to grant that higher
order terms could diverge increasing the number of photons. The Haroche's group
experiments proved definitely that the scenario described by Gea-Banacloche is
essentially correct. The most important point in all the matter is to take the limit
of a large number of photons. This has the effect to push to infinity the
revival time \cite{ebe} leaving just a collapsed wave function for all practical
purposes.

Actually, this problem is double-faced and one may ask if this same effect occurs
when the number of two-level atoms is taken to go to infinity, both with the coupling
with the radiation field being a constant or not as this generally depends on 
$V^{-\frac{1}{2}}$, being $V$ the volume in which the radiation field stays. 
Different limits are proper to different experiments that can
be devised in quantum optics (e.g. ion traps or cavities with constant volume).
The model to be considered in this case is the Dicke model \cite{dic} that has
$N$ two-level atoms interacting with a single radiation mode.
Pioneering work on appearance of classicality in the thermodynamic limit in Dicke model
was carried out by Lieb and Hepp \cite{hl}. 
This model has been extensively studied recently by Emary and Brandes 
\cite{em1,em2} proving the existence of a
quantum phase transition for a given coupling in the thermodynamic limit,
taking both the number of atoms and the volume going to infinity and keeping the
density constant. Overcoming the critical point, parameterized by the coupling
constant, one has a superradiant phase that, if the thermodynamic limit is not
taken, can produce macroscopic superposition states evolving coherently. The point
is that the thermodynamic limit grants that such states disappear, classicality
emerges and the model becomes integrable. In view of the experiments of Haroche's
group, these results are very important giving an analogous conclusion with respect
to the case of photons that classicality can emerge in the thermodynamic limit.

In the Haroche's group experiment the volume of the cavities is kept constant
while the number of photons was varied. This is a rather common situation in quantum
optics as varying a volume can be very difficult to realize and one has to limit
to consider just an increasing number of particles. This situation for the Dicke
model was studied in \cite{fra1,fra2} after a general theory of the emerging of
classicality in the thermodynamic limit was proposed \cite{fra3,fra4b,fra4}. The conclusion
has been the same given in the work of Gea-Banacloche and Emary and Brandes that
increasing the number of particles, superposition states disappear and classicality
emerges. This conclusion can now be rigorously proved by the Holstein-Primakoff
transformation \cite{hp} giving a general result, that could only be supposed in
\cite{fra2}, granting the proper behavior in this limit of the Dicke Hamiltonian
with respect to the measurement problem. The essential point is to change the
quantization axis for the two-level atoms from the Hamiltonian 
of the atoms to the
interaction, reversing the situation with respect to the works of Emary and Brandes

This analysis is fundamental as the electromagnetic interaction is the essential
tool for any measurement apparatus. 
In turn, this means that the environmental decoherence
program described in \cite{zur1} can be completed, pointing toward a solution of
the measurement problem, supporting it with a proper reductionist approach based
on quantum electrodynamics and thermodynamic limit. 
This situation is very similar to the one faced in
statistical mechanics where the emerging of phase transitions was not clear
being the partition function an entire function. 
The answer, after Onsager's solution of the two-dimensional Ising model \cite{ons}, 
was given by Yang and Lee's theorems implying the thermodynamic limit \cite{yl}.

A relevant aspect that our analysis and the one of Emary and Brandes imply is that
a fully polarized state is assumed for the classicality to emerge. Such a situation
has been exploited in mesoscopic physics to approach two open problems observed
in quantum point contacts and in general devices as nanowires and quantum dots
\cite{fra5,fra6,qpc1,qpc2,qpc3} and sometimes connected to appearing of 
classicality where was not expected. In our analysis this aspect will be
just pointed out being already well studied. Anyhow, the dependence on the initial
preparation is crucial and this fact was already pointed out by Zurek \cite{zur2}.

The paper is so structured. In Sec.\ref{sec2} we introduce the Dicke model and 
analyze the Holstein-Primakoff transformation as generally found in literature.
In Sec.\ref{sec3} we point out how, asking for a different limit procedure as
either increasing the number of particle or the coupling constant, a different
view is required, obtaining the correct Holstein-Primakoff transformation in
this case. In Sec.\ref{sec4} an analysis is carried out of some possible applications
of the discussed results. Finally, in Sec.\ref{sec5} the conclusions are given. 

\section{Dicke model and Holstein-Primakoff transformation}
\label{sec2}

The Hamiltonian of the Dicke model is generally written as
\begin{equation}
    H = \frac{\Delta}{2}\sum_{n=1}^N\sigma_{zi}+\omega a^\dagger a +
	g\sum_{n=1}^N\sigma_{xi}(a^\dagger + a)
\end{equation}
being $\Delta$ the separation between the levels of the two-level atoms,
$g$ the coupling constant, $N$ the number of two-level atoms, $a^\dagger$, $a$
the creation and annihilation operators for the radiation mode, $\sigma_{xi}$,
$\sigma_{zi}$ Pauli spin matrices for the i-th atom. The coupling constant
depends on the volume $V$ as $V^{-\frac{1}{2}}$. Introducing the spin operators
\begin{eqnarray}
    S_x &=& \frac{1}{2}\sum_{n=1}^N\sigma_{xi} \\ \nonumber
    S_z &=& \frac{1}{2}\sum_{n=1}^N\sigma_{zi}	
\end{eqnarray}
the Hamiltonian can be cast in the form
\begin{equation}
    H = \Delta S_z + \omega a^\dagger a + 2gS_x(a^\dagger + a).
\end{equation}
This model displays a rich dynamics as shown in the recent papers by Emary and Brandes
\cite{em1,em2}. In fact, the presence of a second order phase transition
was proven and also quantum chaos was shown to appear. These studies are generally
carried out in the thermodynamic limit taking $N\rightarrow\infty$,
$V\rightarrow\infty$ and $\frac{N}{V}$ is kept constant. This can be done by 
introducing the coupling constant $\lambda=\sqrt{N}g$ and rewriting the Hamiltonian in the form
\begin{equation}
\label{eq:hd}
    H = \Delta S_z + \omega a^\dagger a + 2\frac{\lambda}{\sqrt{N}}S_x(a^\dagger + a).
\end{equation}
Choosing as quantization axis $S_z$, we note that $S_x=\frac{1}{2}(S_++S_-)$ so,
finally
\begin{equation}
\label{eq:hbe}
    H = \Delta S_z + \omega a^\dagger a + \frac{\lambda}{\sqrt{N}}(S_++S_-)(a^\dagger + a)
\end{equation}
that is the form considered in refs.\cite{em1,em2}. 

In order to analyze the Hamiltonian (\ref{eq:hbe}) in the thermodynamic limit, one
recurs to the Holstein-Primakoff transformation \cite{hp} that can be written as
\begin{eqnarray}
    S_+ &=& \sqrt{N}b^\dagger\left(1-\frac{b^\dagger b}{N}\right)^{\frac{1}{2}} \\ \nonumber
	S_- &=& \sqrt{N}\left(1-\frac{b^\dagger b}{N}\right)^{\frac{1}{2}}b \\ \nonumber
	S_z &=& b^\dagger b - \frac{N}{2}
\end{eqnarray}
being $b^\dagger$ and $b$ bosonic creation and annihilation operators such that
$[b,b^\dagger]=1$. The Hamiltonian (\ref{eq:hbe}) takes the form
\begin{equation}
    H = -\frac{N}{2}\Delta + \Delta b^\dagger b + \omega a^\dagger a
	+ \lambda\left[
	b^\dagger\left(1-\frac{b^\dagger b}{N}\right)^{\frac{1}{2}} +
	\left(1-\frac{b^\dagger b}{N}\right)^{\frac{1}{2}}b
	\right](a^\dagger + a).
\end{equation}
Now, we are in a position to obtain a series in $\frac{1}{N}$ for the Dicke model
yielding at different orders, omitting the constant,
\begin{eqnarray}
    H_0 &=& \Delta b^\dagger b + \omega a^\dagger a + \lambda(a^\dagger + a)(b^\dagger +b) \\ \nonumber
	H_1 &=& -\frac{\lambda}{2N}(b^\dagger b^\dagger b + b^\dagger bb) \\ \nonumber
	&\ldots&
\end{eqnarray}
It is immediately realized that the leading order Hamiltonian $H_0$, describing the
behavior of the model in the thermodynamic limit, can be straightforwardly diagonalized.
This means that in the thermodynamic limit the Dicke model is exactly integrable.
In refs.\cite{em1,em2} was also shown the appearance of a quantum phase transition
at the critical coupling constant $\lambda_c=\frac{\sqrt{\Delta\omega}}{2}$. These
authors also showed that any macroscopic coherence disappears in the thermodynamic limit
and, by a numerical effort, that, keeping $N$ fixed and letting the coupling
constant increase as $\lambda\rightarrow\infty$, the model is well described by
the Hamiltonian
\begin{equation}
    H = \omega a^\dagger a + \frac{2\lambda}{\sqrt{N}}S_x(a^\dagger + a)
\end{equation}
with the proper quantization axis being given by $S_x$. Our aim, beside the
others, will be to obtain this result analytically in the same context as above,  
even if in refs.\cite{fra3,fra4b} we have derived it by a perturbation series.
Anyhow, the numerical result gives fully support to our preceding work. 
 
\section{The limit of a large number of particles and the strong coupling limit}
\label{sec3}

In experiments in quantum optics it is not generally easy to have the volume varying with
the increasing of the number of particles but two other limits can be interesting as well.
One can consider to increase the number of particles $N\rightarrow\infty$ or the coupling
constant $g\rightarrow\infty$. The former can be also called thermodynamic limit as the
volume entering into the coupling constant could not be generally linked to the 
increasing volume of the particles interacting with the radiation field. 
Both these cases can be treated in the same way as we are going to see but the approach
have to be changed as the proper quantization axis is no more the one of $S_z$ but $S_x$. Again we
start from the Hamiltonian (\ref{eq:hd}) but now, the Holstein-Primakoff transformation must be
\begin{eqnarray}
    S_x &=& -\frac{N}{2}+c^\dagger c \\ \nonumber
	S_+ &=& \sqrt{N}c^\dagger\left(1-\frac{c^\dagger c}{N}\right)^\frac{1}{2} \\ \nonumber
	S_- &=& \sqrt{N}\left(1-\frac{c^\dagger c}{N}\right)^\frac{1}{2}c,
\end{eqnarray}
being $S_z = \frac{1}{2}(S_++S_-)$. These give
\begin{equation}
    H = \frac{\Delta}{2}\sqrt{N}
	\left[c^\dagger\left(1-\frac{c^\dagger c}{N}\right)^\frac{1}{2}+
	\left(1-\frac{c^\dagger c}{N}\right)^\frac{1}{2}c\right]
	+2gc^\dagger c (a+a^\dagger)
	+\omega a^\dagger a-Ng(a+a^\dagger).
\end{equation}
The $\frac{1}{N}$ series is now
\begin{equation}
    H = \omega a^\dagger a-Ng(a+a^\dagger) + 2gc^\dagger c (a+a^\dagger)+
	\frac{\Delta}{2}\sqrt{N}
	(c^\dagger+c)-\frac{\Delta}{4}\frac{1}{\sqrt{N}}(c^\dagger c^\dagger c + c^\dagger cc)+\ldots.
\end{equation}
This Hamiltonian is not as easy to manage unless we apply the unitary transformation
\begin{equation}
    U_0 = e^{\frac{2g}{\omega}c^\dagger c(a-a^\dagger)}
\end{equation}
giving the transformed Hamiltonian
\begin{eqnarray}
    H'&=& \omega a^\dagger a-Ng(a+a^\dagger) + 4N\frac{g^2}{\omega}c^\dagger c
	+ \frac{\Delta}{2}\sqrt{N}\left[c^\dagger e^{-\frac{2g}{\omega}(a-a^\dagger)}
	+c e^{\frac{2g}{\omega}(a-a^\dagger)}\right] \\ \nonumber
	&-& 4\frac{g^2}{\omega}(c^\dagger c)^2 \\ \nonumber
	&-& \frac{\Delta}{4N}\left[c^\dagger c^\dagger c e^{-\frac{2g}{\omega}(a-a^\dagger)}
	+c^\dagger cc e^{\frac{2g}{\omega}(a-a^\dagger)}\right] + \ldots
\end{eqnarray}
and this is a $\frac{1}{N}$ series with
\begin{eqnarray}
    H'_0 &=& \omega a^\dagger a - Ng(a+a^\dagger) + 4N\frac{g^2}{\omega}c^\dagger c \\ \nonumber
	H'_1 &=& \frac{\Delta}{2}\sqrt{N}\left[c^\dagger e^{-\frac{2g}{\omega}(a-a^\dagger)}
	+c e^{\frac{2g}{\omega}(a-a^\dagger)}\right] \\ \nonumber
	H'_2 &=& - 4\frac{g^2}{\omega}(c^\dagger c)^2 \\ \nonumber
	H'_3 &=& - \frac{\Delta}{4N}\left[c^\dagger c^\dagger c e^{-\frac{2g}{\omega}(a-a^\dagger)}
	+c^\dagger cc e^{\frac{2g}{\omega}(a-a^\dagger)}\right] \\ \nonumber
	&\vdots& 
\end{eqnarray}
At this stage we have proved the result that, for $N\rightarrow\infty$, the effective Hamiltonian
to consider is $H'_0$. We now show that this result holds also for $N$ being fixed and 
$g\rightarrow\infty$ as proved numerically in \cite{em2} proving that the same effect can be obtained
in different physical situations. To work out our proof we need the leading order eigenvalues
\begin{equation}
    E_{mn}^{(0)}=m\Omega+n\omega - \frac{N^2g^2}{\omega}
\end{equation}
with $\Omega = 4N\frac{g^2}{\omega}$, and the eigenstates
\begin{equation}
\label{eq:lead}
    |m;n\rangle = |m\rangle_- e^{\frac{g}{\omega}(N-2m)(a^\dagger-a)}|n\rangle_+ 
\end{equation}
and having set $c^\dagger c|m\rangle_-=m|m\rangle_-$ and $a^\dagger a|n\rangle_+=n|n\rangle_+$
the eigenstates of the harmonic oscillator. We can recognize at this stage that the radiation field
is ruled by displaced number states \cite{kn3} while the atoms are ruled by number
states with $m=0$ corresponding to the lowest state with $S_x=-\frac{N}{2}$.

To accomplish our aim, we apply Schr\"odinger-Rayleigh perturbation theory in order to
evaluate higher order corrections. This series is obtained by using the transformed Hamiltonians
and the unperturbed eigenstates $|m\rangle_- e^{\frac{Ng}{\omega}(a^\dagger-a)}|n\rangle_+$. 
It is easy to see that for the first order correction due to $H'_1$ one has
\begin{equation}
    E_{mn}^{(1)} = 0
\end{equation}
and
\begin{equation}
    |m;n\rangle^{(1)} = \frac{\Delta}{2}\sqrt{N}\sum_{n_1\neq n}
	\left[
	\frac{\sqrt{m+1}C_{n,n_1}\left(\frac{2g}{\omega}\right)}{(n-n_1)\omega-\frac{4Ng^2}{\omega}}|m+1;n_1\rangle +
	\frac{\sqrt{m}C_{n,n_1}\left(-\frac{2g}{\omega}\right)}{(n-n_1)\omega+\frac{4Ng^2}{\omega}}|m-1;n_1\rangle
	\right]
\end{equation}
having set
\begin{equation}
    C_{n,n_1}\left(\pm\frac{2g}{\omega}\right) = \ _+\langle n_1|e^{\pm\frac{2g}{\omega}(a^\dagger-a)}|n\rangle_+.
\end{equation}
The next correction is computed by $H'_2$ that, making the same computation as above, yields
\begin{equation}
    E_m^{(2)} = -4\frac{g^2}{\omega}m^2
\end{equation}
for the eigenvalues and
\begin{equation}
    |m;n\rangle^{(2)} = 0
\end{equation}
for the eigenstates. Using $H'_3$ gives
\begin{equation}
    E_m^{(3)} = 0
\end{equation}
for the eigenvalues and
\begin{equation}
    |m;n\rangle^{(3)} = -\frac{\Delta}{4\sqrt{N}}\sum_{n_1\neq n}
	\left[
	\frac{m\sqrt{m+1}C_{n,n_1}\left(\frac{2g}{\omega}\right)}{(n-n_1)\omega-\frac{4Ng^2}{\omega}}|m+1;n_1\rangle +
	\frac{(m-1)\sqrt{m}C_{n,n_1}\left(-\frac{2g}{\omega}\right)}{(n-n_1)\omega+\frac{4Ng^2}{\omega}}|m-1;n_1\rangle
	\right]
\end{equation}
for the eigenstates.
One can go on with this computation to first order in the Schr\"odinger-Rayleigh expansion and we
can see some regularities. One can see that even terms produce corrections to the eigenvalues
while a odd terms produce corrections the eigenstates. Besides, this expansion
has the structure of a $\frac{1}{\sqrt{N}}$ as already pointed out in \cite{fra2}. We also observe
that this expansion is also meaningful when $N$ is kept at a fixed value and the strong coupling
limit $g\rightarrow\infty$ is taken, in agreement with the numerical result presented in \cite{em2}.
The fundamental result that can be drawn from this computation is that in the limit
of $N\rightarrow\infty$ or $g\rightarrow\infty$ or both, the leading order
solution (\ref{eq:lead}) becomes exact. This proves our guess in ref.\cite{fra2} and fully support
the conclusions given there. Finally, we note that the above expansion is not meaningful in the
case $N\rightarrow\infty$, $g\rightarrow 0$ and $\sqrt{N}g=constant$ that
was studied in ref.(\cite{em1,em2}). In this latter case the proper quantization axis is given by
$S_z$.

For our aims it is also important to obtain the time evolution unitary operator
at the leading order. One has
\begin{equation}
\label{eq:ud}
    U(t) = e^{i\frac{N^2g^2}{\omega}t}\sum_{m,n} e^{-i(m\Omega+n\omega)t}|m;n\rangle\langle m;n|.
\end{equation}
The most interesting results are obtained with this time evolution when the ensemble of two-level atoms
is in the ground state as we will see in the next section. 

\section{Experimental implications}
\label{sec4}

The properties of the Dicke model in the thermodynamic limit can have important experimental
consequences. Besides, being by itself a meaningful model in quantum electrodynamics, it can
also have impact on the problem of measurement, as already said, because electromagnetic
interactions represent the main tool to operate measurements on a quantum system. As a matter
of fact, this approach can represent a conceptual completion to the decoherence program. The
recent experiments by Haroche's group \cite{har1,har2} heavily point out that classicality
may be an emerging property of quantum systems in the thermodynamic limit.

On this line of thought, two relevant effects have been pointed out by us for the Dicke model
\cite{fra1,fra2,fra3}. In any case we assume that the ensemble of two-level atoms is in the ground
state, a kind of ``ferromagnetic state'' that we proved to produce decoherence \cite{fra5,fra6}
in a different context.

A first interesting case is given by assuming the radiation field in the ground state as is
the ensemble of two-level atoms, so $|\psi(0)\rangle = |0\rangle_+|0\rangle_-$. 
Using the unitary evolution given in eq.(\ref{eq:ud}) one has
\begin{equation}
    |\psi(t)\rangle = e^{i\frac{N^2g^2}{\omega}t}
	\sum_{n} e^{-in\omega t}\ _+\langle n|e^{\frac{Ng}{\omega}(a^\dagger-a)}|0\rangle_+
	|0;n\rangle.
\end{equation}
Using the relation
\begin{equation}
    \ _+\langle n|e^{\frac{Ng}{\omega}(a^\dagger-a)}|0\rangle_+ = 
	e^{-\frac{N^2g^2}{\omega^2}}\left(\frac{Ng}{\omega}\right)^n\frac{1}{\sqrt{n!}}
\end{equation}
it is straightforward to obtain the result already given in ref.\cite{fra3}
\begin{equation}
    |\psi(t)\rangle = |0\rangle_- e^{i\frac{N^2g^2}{\omega^2}(\omega t-\sin(\omega t))}
	|\beta(t)\rangle
\end{equation}
being $|\beta(t)\rangle$ a coherent state with 
$\beta(t) = \frac{Ng}{\omega}(1-e^{i\omega t})$. It is not difficult to see that here we have
a quantum amplifier (QAMP): We started with both the ensemble of two-level atoms and the
radiation field in their ground states and the time evolution, in the proper limits of 
$N\rightarrow\infty$ and $g$ fixed or $g\rightarrow\infty$ and $N$ fixed, produces a
macroscopic radiation state while the state of the ensemble of two-level atoms is left
untouched, acting as a passive amplifying medium. This amplification effect, that we have
derived in different ways, can be very useful once it is considered in the framework of
theory of measurement in quantum mechanics. The main point to rely on is the fact that the
principal mechanism of interaction between a measurement apparatus and a quantum system is
the electromagnetic interaction. Having uncovered the above effect in the thermodynamic limit
permits us to implement in the decoherence program a reductionist approach.

The next effect to be considered for the Dicke Hamiltonian in the thermodynamic limit is
a ``collapse of the wave function''. Normally, one assumes that, at a classical level, no
superposition states of macroscopic objects are ever observed. This in turn means that one
should expect the disappearance of such states wherever they can form. So, let us
consider a superposition of two coherent states as
\begin{equation}
    |\chi(0)\rangle = {\cal N} (|\gamma e^{i\phi}\rangle + |\gamma e^{-i\phi}\rangle)
\end{equation}
being $\cal N$ a normalization factor, $\gamma$ and $\phi$ two real parameters. This state can
be termed a phase Schr\"odinger cat state. As before, we assume that the ensemble of two-level
atoms is in the ground state. Using eq.(\ref{eq:ud}) one has \cite{fra1,fra2}
\begin{equation}
     |\psi(t)\rangle = e^{i\xi(t)}{\cal N}(e^{i\phi_1(t)}
	 |\frac{Ng}{\omega}(e^{i\omega t}-1) 
	 + \gamma e^{i\phi-i\omega t}\rangle 
	 + e^{i\phi_2(t)}|\frac{Ng}{\omega}(e^{i\omega t}-1) + \gamma e^{-i\phi-i\omega t}\rangle)|0\rangle_-
\end{equation}
being
\begin{equation}
    \xi(t)=\frac{N^2g^2}{\omega^2}(\omega t - \sin(\omega t))
\end{equation}
and
\begin{eqnarray}
    \phi_1(t) &=& -\gamma\frac{Ng}{\omega}[\sin\phi+\sin(\omega t -\phi)] \\ \nonumber
    \phi_2(t) &=&\gamma\frac{Ng}{\omega}[\sin\phi-\sin(\omega t +\phi)].
\end{eqnarray}
The result we obtain is rather surprising as we easily realize that when the limits $N\rightarrow\infty$
or $g\rightarrow\infty$ or both are taken, the result is a single macroscopic coherent radiation
state and the superposition state disappears. Again, the ensemble of two-level atoms is left untouched
behaving passively. Then, the Dicke Hamiltonian in the thermodynamic limit is able to remove
macroscopic superposition states of radiation fields. The same can be proven to be true also
for a superposition of Fock number states \cite{fra2}. 

From the above analysis by the Holstein-Primakoff transformation, 
we are able to assume that these results hold true at any order and become exact in the
thermodynamic limit. Together with the analysis of Emary and Brandes \cite{em1,em2} we can
conclude that the Dicke model removes quantum coherence in the thermodynamic limit, proving
itself to be the first realistic model of decoherence acting by unitary evolution. 
It is important to emphasize that what makes the argument work is the thermodynamic limit
and the proper choice of the initial state.

\section{Conclusions}
\label{sec5}

Several toy models have been devised so far to understand the working of decoherence but a
proper analysis of quantum electrodynamics in the thermodynamic limit 
has not been carried out till now. But the problem
of measurement theory in quantum mechanics cannot be addressed unless a proper understanding
of radiation-matter interaction is at hand for macroscopic objects. The reason to work out
this understanding is linked to the obvious fact that electromagnetic interaction is the main
tool we have to operate measurements on a quantum system.

On this line of reasoning we have given a complete analysis for a simple but realistic quantum
electrodynamics model that has applications in several experimental situations in cavities
or ion traps. We have shown how macroscopic coherence gets lost in the thermodynamic limit
producing disappearance of superposition states by unitary evolution. Besides, we have
observed that amplification of vacuum fluctuations of a radiation field can be obtained producing
a classical electromagnetic field (QAMP).

The relevance for the decoherence is that a possibility of a completion program for
measurement theory can be at hand with a proper reductionist approach.

In any case, after the experiments by Haroche's group \cite{har1,har2} that support the
mathematical analysis of Gea-Banacloche \cite{gb1,gb2} of the Jaynes-Cummings model in the
limit of a large number of photons, it would be interesting to have a clear understanding, from
an experimental point of view, of what is going on increasing the number of atoms.  


\end{document}